\pdfoutput=1

\newcommand{\eqnref}[1]{Eq.\ (\ref{#1})}

\newcommand{\Secref}[1]{Section \ref{#1}}

\newcommand{\widesubfigsize}[0]{0.68\columnwidth}

\newcommand{\myfig}[5]{
\begin{figure}[{#5}]
\sidecaption
\includegraphics[keepaspectratio,width=#4,angle=0]{#1}%
\caption{#2}\label{#3}%
\end{figure}}

\documentclass[graybox]{sv}

\usepackage{mathptmx}       
\usepackage{helvet}         
\usepackage{courier}        
\usepackage{type1cm}        
\usepackage{makeidx}         
\usepackage{graphicx}        
\usepackage{multicol}        
\usepackage[bottom]{footmisc}

\usepackage{subfigure}        
\usepackage{verbatim}        

\makeindex             

\begin{document}

\title*{Inference of hidden structures in complex physical systems by
multi-scale clustering}


\author{Z. Nussinov*, P. Ronhovde*, Dandan Hu*, S. Chakrabarty, M. Sahu, Bo Sun, N. A. Mauro, and K. K. Sahu}
\institute{Zohar Nussinov, Washington University in St. Louis, MO 63130, USA  \email{zohar@wuphys.wustl.edu} and Department of Condensed Matter Physics, Weizmann Institute of Science, Rehovot 76100, Israel,
Peter Ronhovde, Findlay University, Findlay, OH 45840, USA  \email{ronhovde@findlay.edu},
Dandan Hu, \email{dan1226@gmail.com}, Saurish Chakrabarty, 
Department of Physics, Indian Institute of Science, Bangalore 560012, India \email{schakrab@go.wustl.edu}, Bo Sun, Washington University in St. Louis, MO 63130, USA  \email{bosun@wustl.edu}
Nicholas A. Mauro, 
North Central College, Naperville, Il 60540, \email{Nicholas.mauro@gmail.com},
Kisor K. Sahu, School of Minerals, Metallurgical and Materials Engineering, Indian Institute of Technology, Bhubaneswar-751007, India, \email{kis.sahu@gmail.com}
}
%
%
\maketitle

\abstract{We survey the application of a relatively new branch of statistical physics---``community detection''-- to data mining. In particular, we focus on the diagnosis of materials and automated image segmentation. Community detection describes the quest of partitioning a complex system involving many elements into optimally decoupled subsets or communities of such elements. We review a multiresolution variant which is used to ascertain structures at different spatial and temporal scales. Significant patterns are obtained by examining the correlations between different independent solvers. Similar to other combinatorial optimization problems in the NP complexity class, community detection exhibits several phases. Typically, illuminating orders are revealed by choosing parameters that lead to extremal information theory correlations.}

\section{The general problem}
\label{sec:1}

A basic question that we wish to discuss in this work is whether machine learning and data mining tools may be applied to the analysis of material properties. Specifically, we will review initial efforts to detect, via statistical mechanics and the tools of information science and network analysis, pertinent structures on all scales in general complex systems. We will describe mapping atomic and other configurations onto graphs. As we will explain, patterns found in these graphs via statistical physics methods may inform us about the structure of the investigated materials. These structures can appear on multiple spatial and temporal scales. In comparison to standard procedures, the advantage of such an approach may be significant. 

There are numerous classes of complex systems. One prototypical variety is that of glass forming liquids. ``Glasses'' have been analyzed with disparate tools \cite{angell,fred',nakamura,saida,sordelet,wang,mcgreevy,keen,Sheng,ref:finney,HA,BO,rfot,ref:lubchenkowolynes,gilles,nab}.  Although they have been known for millennia, structural glasses still remain ill understood. It is just over eighty years since the publication of one of the most famous papers concerning the structure of glasses \cite{fred'}. Much has been learned since the early days of hand-built plastic models and drawings, yet basic questions persist.

Amorphous systems such as glasses strongly contrast with idealized simple solids. In simple crystals, the structure of an atomic unit cell is replicated to span the entire system. Long before scattering and tunneling technologies, prominent figures such as Robert Hooke, Christiaan Huygens, and their contemporaries in the 17th century proposed the existence of sharp facets in single crystals results from recurrent fundamental unit cell configurations. The many years since have seen numerous breakthroughs (including the advent of quantum mechanics and atomic physics) and witnessed a remarkable understanding as to how the quintessential simple periodic structure of crystals accounts for many of their properties. However, while simple solids form a fundamental pillar of current technology (e.g., the transistor whose invention was made possible by an understanding of the electronic properties of nicely ordered periodic crystals and chemical substitution therein), there are many other complex systems whose understanding is extremely important yet still lacking. The discovery of salient features of these materials across all scales is important for both applied and basic science. The recognized significance of this problem engendered the {\it Materials Genome Initiative} \cite{wh}---a broad effort to develop infrastructure for accelerating materials innovation.  

This work discusses a path towards solving this problem in complex amorphous materials. The framework that we will principally suggest is that of {\em multi scale community detection}. This approach does not invoke assumptions as to which system properties are important and construct resulting minimal toy models based on the assumptions. The insightful guess-work that is typically required to describe complex materials is, in the work that we review, replaced by a computerized variant of the {\it wisdom of the crowds} phenomena \cite{wisdom}. The key concepts underlying this approach may be applied to general hard problems beyond those concerning the structure of materials or even general data mining. In the next section, we review an ``Information theoretic ensemble minimization'' method that may be suited for such tasks.

\begin{figure}
\sidecaption
\subfigure[\ Independent solvers (or ``replicas'') on a schematic energy landscape.]
  {\includegraphics[width=\widesubfigsize]{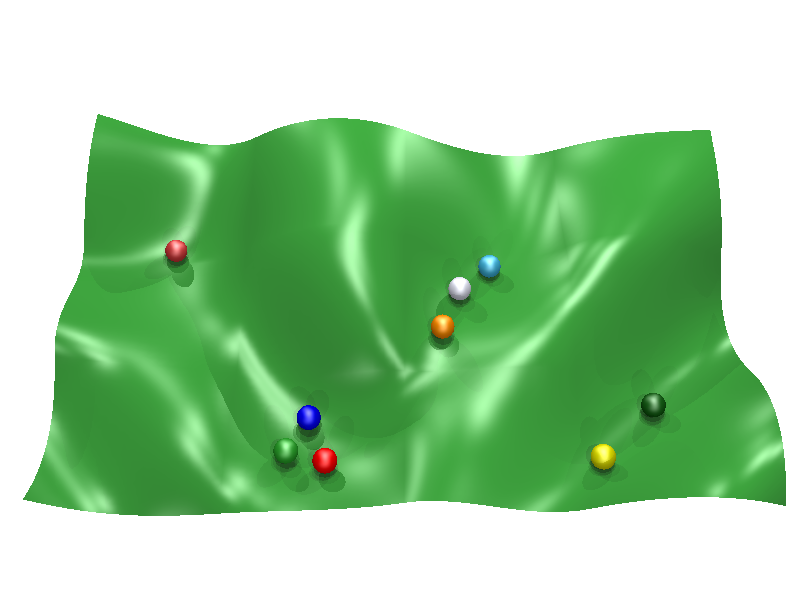}}
\subfigure[\ Coupled solvers.]
  {\includegraphics[width=\widesubfigsize]{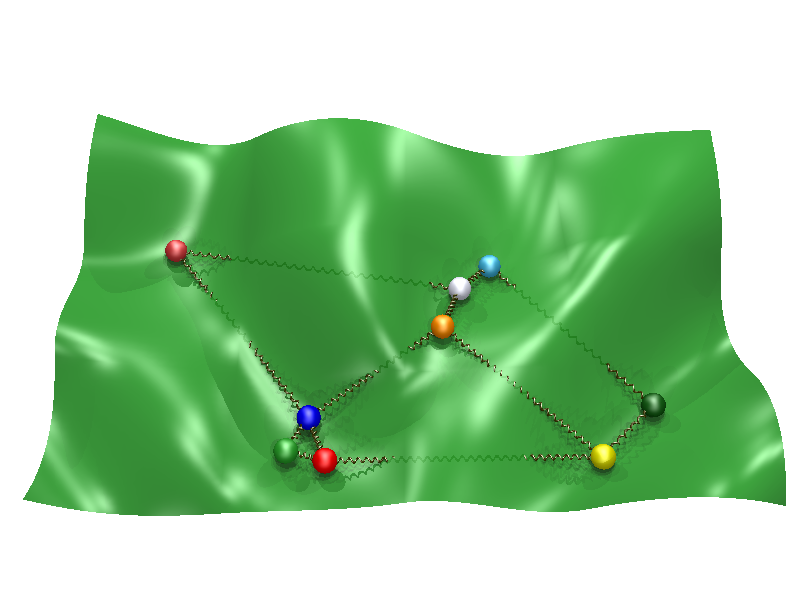}}
\caption{The spheres in panel (a) of the figure depict 
solvers (or ``replicas'') independently navigating the energy landscape defined by \eqnref{APM'}.
Strong correlations among the replicas indicate stable, well-defined partition.
We evaluate agreement among all replica pairs using the information correlations (\Secref{multi}).
In panel (b), interactions between the replicas assist the ensemble in
finding optimal low energy states.}
\label{fig:MRAlandscape}
\end{figure}

\begin{figure}
\sidecaption
\includegraphics[width=0.6\columnwidth]{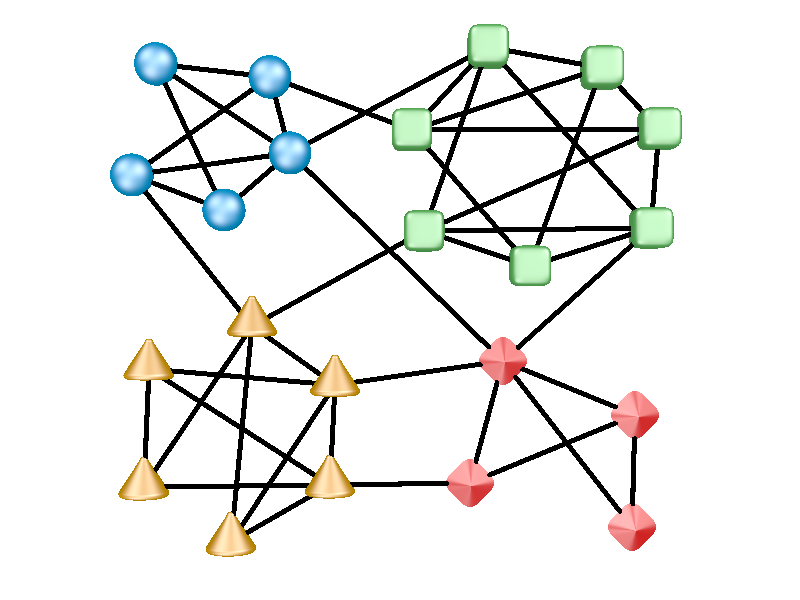}
\caption{A small network partition where
individual communities are represented by different node shapes and colors.
``Friendly'' or ``cooperative'' relations are depicted by solid, black lines.
These are modeled as ferromagnetic interactions in \eqnref{APM'}. 
``Missing'' or ``undefined'' relations work to break up well-defined communities,
so they are modeled with anti-ferromagnetic interactions,
meaning they are repulsive in terms of their energy contributions.
The physical energy model trivially extends to more general relations including 
weighted and adversarial relations (not depicted here).}
\label{fig:simplenetwork}
\end{figure}

\begin{figure}
\sidecaption
\includegraphics[width=0.6\columnwidth]{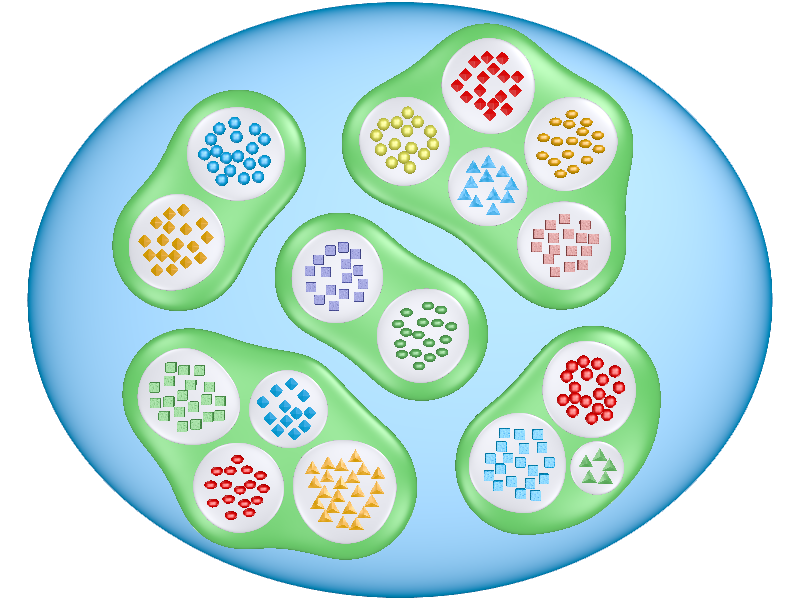}
\caption{A partition of a synthetic network with $256$ nodes having three resolution levels \cite{ref:rzlocal}. 
The random edge density (fraction of edges connecting pairs of points in different communities) is 
10\% on the global scale. At increasing resolution there are five groups with an inter-community edge density of 30\%.
At the highest resolution, these five groups are further split into small sub clusters (16 in total) each having an internal edge density
of 90\%. As described in \Secref{multi}, a multi-resolution algorithm may identify 
different categories of partitions in hierarchical systems.
See Figure \ref{fig:examplehierarchydata} for a demonstration of how the multiresolution algorithm
accurately isolates both levels of the hierarchy.}
\label{fig:examplehierarchy}
\end{figure}

\begin{figure}
\sidecaption
\includegraphics[width=0.6\columnwidth]{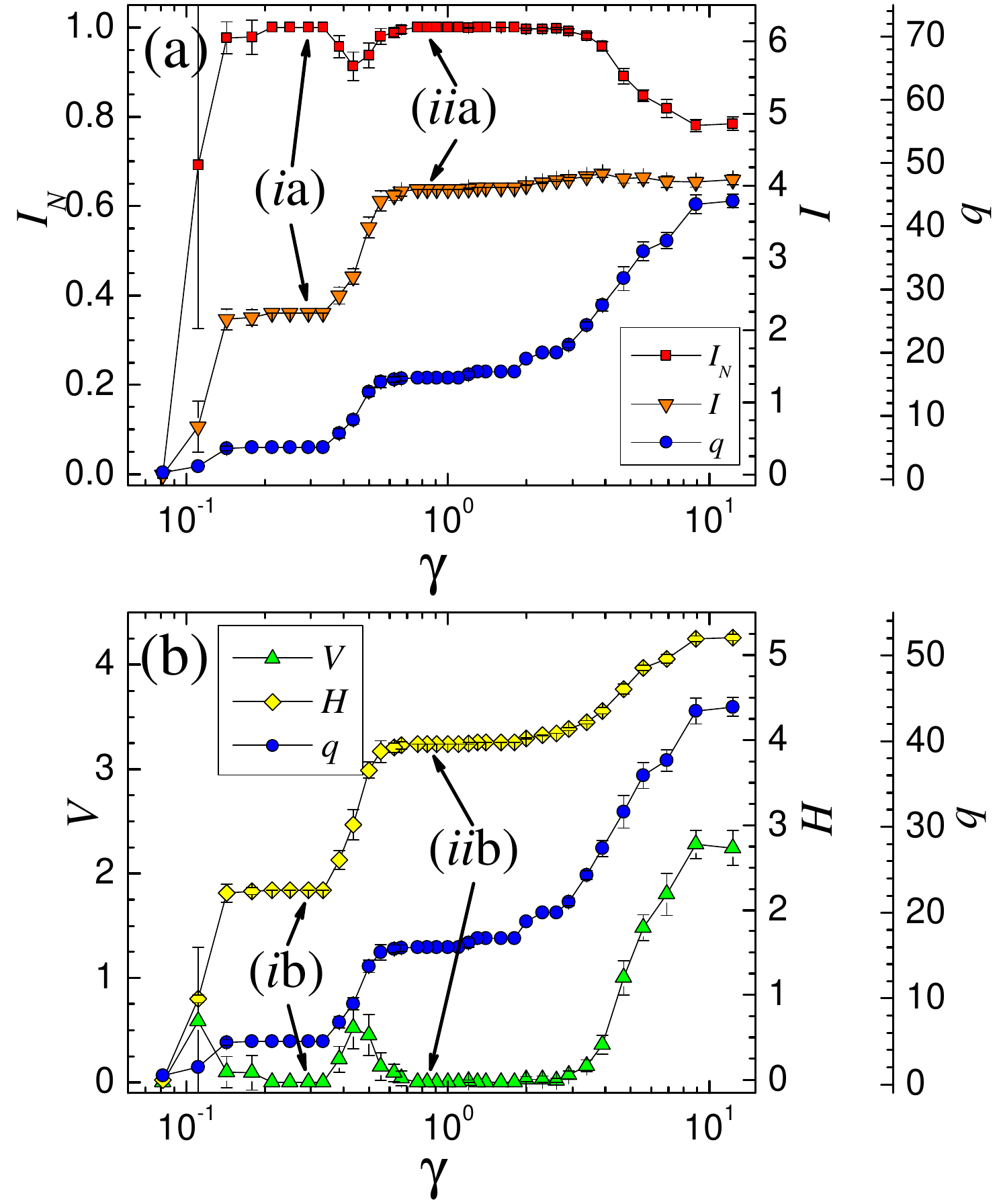}
\caption{Information theoretic and other metrics of the multiresolution algorithm in \Secref{multi}
as applied to the synthetic partition depicted in Figure \ref{fig:examplehierarchy} \cite{ref:rzmultires}. 
In the top panel, the average inter-replica normalized mutual information ($I_N$), (un-normalized) mutual information ($I$), and number of clusters (or communities) $q$ are plotted as a function of the resolution parameter $\gamma$. In the bottom panel, the Shannon entropy ($H$) and the average inter-replica variation of information  ($V$) are further provided. 
As described in the text, stable partitions lead to plateaux (or more general local extrema) in the inter-replica information theory and other correlations as a function
of the resolution parameter. Two such candidate resolutions (marked $(i)$ and $(ii)$) are seen in both panels (a) and (b).  These plateaux show how the multiresolution algorithm
may isolate both level $2$ (superclusters) and level $3$ (smallest clusters) of the hierarchy of Figure \ref{fig:examplehierarchy}.}
\label{fig:examplehierarchydata}
\end{figure}

\section{Ensemble minimization}
\label{ensemble}
Before delving into complex material and network analysis, we first discuss a general strategy for solving hard problems. The concept underlying this approach is perhaps best conveyed by a simple cartoon such as that sketched in Figure \ref{fig:MRAlandscape}(a). In this illustration, each sphere corresponds to an individual solver (or ``replica'') that explores an energy landscape.  On its own, each such sphere might get stuck in a local energy minimum. The collective {\it ensemble} of solvers may, however, thwart such situations more readily as compared to the same single solver algorithm \cite{grant34}. In Figure \ref{fig:MRAlandscape}(b), the individual solvers not only roam the energy landscape but also interact amongst themselves as schematically denoted by springs. If a single solver gets stuck in a false minimum, the other solvers may ``pull it out'' and explore broader regions of the energy landscape.

This collective evolution of individual solvers is quite natural and has appeared in different guises across many fields. It anthropological contexts, this basic principle is known as ``wisdom of the crowds'' \cite{wisdom}. That is, the crowd or ensemble of individuals might do far better than a single solver. Unlike ensemble related approaches such as swarm intelligence \cite{ant} or genetic \cite{genetic} algorithms, relevant problems in our context do not focus exclusively on minimizing a given energy function. Rather, we will try to maximize information theory correlations [the effect of the springs in Figure \ref{fig:MRAlandscape}(b)] while simultaneously minimizing a cost function \cite{ref:rzmultires}. If all (or many) solvers agree on a particular candidate solution then that solution may naturally arise in many instances and may be of the high importance regardless of whether or not it is the absolute minimum of the energy. In the physical problems that we will consider---that of finding natural structures in materials---these considerations are pertinent.

The above discussion is admittedly abstract and may, in principle, pertain to any general problem. We next briefly explain the basic mathematical framework---the {\it community detection} problem---in which we will later couch the material structure detection endeavor.

\section{Community detection and data mining}
\label{CD-section} 
Community detection pertains to the quest of partitioning a given graph or network into its optimally decoupled subgraphs (or so-called communities), e.g.,\cite{fortunato2,pnasn,newman_girvan,fortunato3,fortunato1,blondel,newman_fast,gudkov,RosB,book_comm,pre3,spec,spec',darst'}.  As the reader may anticipate, given the omnipresence of networks and the generality of this task, this problem appears in disparate arenas including biological systems, computer science, homeland security, and countless others. In what follows, we introduce some of the key elements of community detection. The graphs of interest will be composed of nodes where a {\it node} is a fundamental element of an abstracted graph. An {\it edge} in the graph is a defined relationship between two nodes. Edges may be weighted or unweighted where the unweighted case is the one most commonly examined. In our applications, we will need to assign weights to the edges in the graph as we will describe. Similarly, in general applications, edges may be either symmetric or directed.

Now we come to a basic ingredient of community detection. A community corresponds to a subset of nodes that are more cohesively linked (or densely connected for unweighted edges) within their own community than they are to other communities. The above definition might seem a bit loose. Indeed, there are numerous formulations of community detection in the literature. As intuitively one may expect, most of these do, more or less, the same thing. When clear community detection solutions exist, all algorithms quantify the structure of large complex networks in terms of the smaller number of the natural cohesive components. Rather general data structures may be cast in terms of abstract networks. Thus, the community detection problem and other network analysis methods can have direct implications across multiple fields. Indeed, we will elaborate how this occurs for image segmentation and material analysis.

In what follows we will briefly review the rudiments of an ``Absolute Potts Model'' method for community detection \cite{ref:rzlocal} that avoids 
a ``resolution limit'' that an insightful earlier Potts model \cite{RB} exhibited. To cast things generally, we make a simple observation underlying the ``Potts'' characterization. Any partition of the numbered nodes $i=1,2,3, \cdots, N$ into $q$ different communities (the ultimate objective of any community detection algorithm) is an assignment $i \to \sigma_{i}$ where the integer $1 \le \sigma_{i} \le q$ denotes the community number to which node $i$ belongs. With a characterization $\{ \sigma_{i} \}$ in hand, we next construct an energy functional.

To illustrate the basic premise, we first consider an unweighted graph---one in which the link strength $A_{ij}$ between the two nodes $i$ and $j$ is $A_{ij} =1$ if an edge is present between the two nodes and $A_{ij}=0$ if there is no link.  As Figure \ref{fig:simplenetwork} demonstrates, for each pair of nodes there are four principal cases to consider. That is, either (i) the two nodes belong to the same community and have an ``attraction'' between them (i.e., $A_{ij} =1$), (ii) two nodes in the same community can have a missing link between them ($A_{ij} =0$), (iii) the two nodes may belong to different communities yet nevertheless exhibit cohesion between themselves ($A_{ij} =1$), or (iv) nodes $i$ and $j$ may belong to different communities and have no edge connecting them ($A_{ij} =0$).  Situations (i) and (iv) agree with the intuitive expectation that nodes in the same community should be connected to one another while those in different communities ought to be disjoint. We may take these four possibilities as the foundation of an energy function. That is, any given pair of nodes may be examined to see which of these categories it belongs to. Thus, a contending cost function is given by the Potts model Hamiltonian
\begin{eqnarray}
\label{APM}
H = - \frac{1}{2} \sum_{i \neq j} [A_{ij} \delta (\sigma_{i}, \sigma_{j}) + \gamma (1-A_{ij}) (1- \delta(\sigma_{i}, \sigma_{j}))].
\end{eqnarray}
In Eq. (\ref{APM}), $\delta(\sigma_{i},\sigma_{j})$ is a Kroncker delta (i.e., $\delta(\sigma_{i}= \sigma_{j}) =1$, $\delta(\sigma_{i} \neq \sigma_{j}) =0$) and $\gamma$ is a ``resolution parameter'' that will play a notable role in our analysis. 
Before turning to the origin of the name of this parameter, we observe that, subtracting an innocuous additive constant, Eq. (\ref{APM}) is trivially
\begin{eqnarray}
H = - \frac{1}{2} \sum_{i \neq j} [A_{ij} -  \gamma (1-A_{ij})] \delta(\sigma_{i},\sigma_{j}). 
\label{APM'}
\end{eqnarray}
As Eq. (\ref{APM'}) makes clear, by virtue of the Kronecker delta $\delta(\sigma_{i},\sigma_{j})$, the sum is {\it local}---i.e., the sum only includes intra-community node pairs. The Hamiltonian of Eq. (\ref{APM'}) may be minimized by a host of methods. In practice, when the solution of the problem is easy to find, nearly all viable approaches will yield the same answer. Amongst many others, two approaches are afforded by spectral methods [in which the discrete Potts model spins are effectively replaced by continuous spherical model (or large $n$) spins] and a conceptually more primitive steepest descent type approach. 

A simple incarnation of the relatively successful greedy algorithm \cite{ref:rzlocal,ref:rzmultires} that extends certain ideas introduced in \cite{blondel} is given by the following steps: (a) Initially, each node forms its own community [i.e., if there are $N$ (numbered) nodes then there will be $q=N$ communities]. 

(b) A node (whose number is $i_1$) is chosen stochastically and then another edge sharing node $i'$ is picked at random.  (c) If it is energetically profitable to move the node $i'$ together into the group formed by $i_1$ then this is done (otherwise community assignments are unchanged). (d) Yet another node $i_2$ is next chosen and once again it is asked whether moving yet another node into the community of $i_2$ lowers the energy. As earlier mentioned, if this change lowers the energy of Eq. (\ref{APM'}), the nodes will be merged. Otherwise no change will be made. (e) In this manner, we cycle through each of the $N$ nodes and repeat as necessary. (f) The process stops and a candidate partition is found once all further possible mergers do not lower the energy further.
As the reader can appreciate, such a simple simple algorithm lowers the energy until the system becomes trapped in a local minimum. 
To improve the accuracy (i.e., further lower the energy of candidate solutions), one may repeat the above steps a finite number of times for a finite number of {\it trials}---i.e., repeat the above when vertices $i_1,i_2, \cdots, i_N$ are chosen in a different random order to see if a lower energy solution may result. 

For the wide range of examined problems, the number of trials for each replica of the system is typically on the order of ten or smaller. When approaching the ``hard phase'' (to be discussed in Section \ref{CDPD}) with multiple false minima, an increase in the number of trials may likely further increase the accuracy (this rise in the accuracy
was termed the ``computational susceptibility'' in \cite{ref:rzmultires,phases}). Typically, elsewhere the improvement in the precision due to a further increase in the number of trials is nearly nonexistent
(see, e.g., Figure 13 in \cite{ref:rzmultires}). Further embellishments of the bare algorithm outlined above, 
include the acceptance of zero energy moves and other refinements \cite{ref:rzlocal}. Other illuminating greedy type approaches for the inference of community structure have been advanced, e.g., \cite{greed}.

\section{Multi-scale community detection}
\label{multi}
We now turn to ``multi-scale'' community detection, e.g., \cite{ref:rzmultires,KSKK,local-multi,jeub,manilo,pt,eytan}. In certain notable approaches, e.g., \cite{eytan}, detection of scale is performed without the resolution parameter but rather by examining the effects of thermal fluctuations in a pure ferromagnetic system (one sans the antiferromagnetic
interaction present in the second term of Eq. (\ref{APM'})), and other considerations elsewhere. In what follows, we will build on the ideas introduced in 
Section \ref{CD-section} that lead to an accurate determination of structure on diverse pertinent scales. To understand the physical content of the resolution parameter (and the origin of its name) in Eq. (\ref{APM'}), we consider several trivial limits.  First, we focus on the case of $\gamma =0$. 
In such a situation, the energy of Eq. (\ref{APM'}) is minimized when all nodes belong to a single community. This is the lowest energy solution since each intra-community link lowers the energy [the first term of Eq. (\ref{APM'})], but there is no energy penalty from any missing links between nodes in the same community since the second term in Eq. (\ref{APM'}) is trivially zero. 
Thus, in order to maximize the number of internal links it is profitable to assign all nodes to the same community. In the diametrically opposite limit---that of $\gamma \to \infty$, the energy penalty diverges unless every pair of nodes belonging to the same community share a link. Thus, in this limit, the lowest energy states are those in which the the system fragments into (typically) a large number of communities where each node is connected to all other nodes in its community. That is, the communities are ``perfect cliques.'' 
As $\gamma$ is monotonically increased from zero, the ground states of Eq. (\ref{APM'}) lead to communities that veer from the extreme global case ($\gamma =0$) to the limit of many disparate densely internally connected local communities ($\gamma \to \infty$). Putting all of the pieces together, the reader can see why $\gamma$ is inherently related to the intra-community edge density and thus is indeed a ``resolution parameter''.  

At this stage, it is not yet clear which values $\gamma$ should be assigned in order to lead to the most physically pertinent solutions.
The non-uniqueness of $\gamma$ is, actually, a virtue of the Potts model based approach of Eq. (\ref{APM'}). That is, in general, there may be several relevant resolution scales that lead to different insightful candidate low energy partitions of this Hamiltonian. This is the situation which is schematically depicted in Figure \ref{fig:examplehierarchy} for a synthetic system that exhibits a hierarchical structure. In such cases as $\gamma$ is increased, the minima of Eq. (\ref{APM'}) unveil different resolutions in the hierarchy.  In practice, the multi-resolution community-detection method \cite{ref:rzmultires} systematically infers the pertinent scale(s) by information-theory-based correlations \cite{info11,info2,info3,ginestra} between different independent solvers (or ``replicas'', as discussed in Section \ref{ensemble}) of the same community detection problem.  In most studied systems, the number of replicas used is $s \le 12$. As alluded to in Section \ref{CD-section}, the lowest energy solution amongst a fixed number of trials is taken for each of the individual replicas. 
If these solvers (i.e., the replicas) strongly concur with each other about local or global features of the solution \cite{ref:rzmultires}, then these aspects are likely to be correct. Such an agreement between solvers is manifest in the information correlations. Information theory extrema \cite{grant3,grant7,grant6} then provide \emph{all relevant system scales}. 

Figure \ref{fig:examplehierarchydata} shows the results of our analysis as the resolution parameter $\gamma$ is varied for the synthetic system of Figure \ref{fig:examplehierarchy}. Plotted are three information theory correlations between replicas---the average inter-replica variation of information (VI), the mutual information (I), the normalized mutual information (NMI), the total number of communities ($q$) found for different values of $\gamma$, and the Shannon entropy ($H$) averaged over different replicas. Transitions between viable solutions are evident as jumps in the number of communities $q$ and, most notably, as transitions between crisp information theory measure plateaux. As shown, each of the plateaux in Figure \ref{fig:examplehierarchydata} corresponds to a different level of the hierarchy of the synthetic network in Figure \ref{fig:examplehierarchy}. Similar to our discussion in Section \ref{CD-section}, in practice the replicas differ from one another in the order in which consecutive vertices are picked and moved so as to minimize the energy of Eq. (\ref{APM'}). Thus, for any given problem has an ensemble of very similar (or nearly identical) viable solutions associated with it. A detailed summary of this approach appears in \cite{ref:rzmultires}.

In accord with the above explanation, as $\gamma$ is increased, the associated candidate energy minima partition the system into more local, smaller communities (deeper levels of the hierarchy). The inter-replica information theory correlations further afford a measure of the quality of the viable partitions. High NMI values (i.e., of size close to unity) indicate solutions that are likely to be pertinent. In the spirit of Section \ref{ensemble}, if the different replicas all agree with one another on a putative partition, then that partition is likely to be physically meaningful. The variation of information measures the disparity between candidate solutions; thus the VI values are high between different NMI plateaux and are low within the NMI plateaux. 

\section{Image segmentation}
\label{image}
Our goal is to identify structure in materials, but before turning to this endeavor, we first illustrate how patterns may, literally, be revealed by community detection. The ideas underlying this objective will elucidate our approach to material genomics. The aim of image segmentation \cite{grant6,book-image,Shi,ling,abin,grant21,grant21.5} is to divide a given digital image into separate objects (or segments) based on visual characteristics. Two somewhat challenging examples are provided in Figure (\ref{fig:hardzebradog}) \cite{chal1,chal2}.

To transform the problem into that of community detection, we map a digital image into a network as follows. Each pixel in an image is regarded as a node in a graph.  (2) The edge weights between nodes in the graph are determined by the degree of similarity between the additive color RGB (i.e., the Red, Green, and Blue) strength of individual pixels or, more generally, of finite size boxes geometrically centered about a given pixel. The bare edge strengths may be embellished and replaced by weights set by the Fourier weights associated with finite size blocks about a given node. Alternatively, we can use exponential weighting of the inter-node edge strength based on the geometric distance between them (the distance between the centers of the finite size blocks about them) \cite{grant6}. The edge value assignment is such that if two pixels $i$ and $j$ (or boxes centered about them) have similar RGB values (or absolute Fourier magnitudes), then a function $V_{ij}$ set by these differences will be small. Analogously, if nodes $i$ and $j$ (or boxes centered around them) are dissimilar then $V_{ij}$ will become large. 

With such functions $V_{ij}$ at hand, a simple generalization of Eq. (\ref{APM'}) is given by
\begin{eqnarray}
\label{weigh}
H = \frac{1}{2}  \sum_{s=1}^{q} \sum_{i,j \in C_{s}} \Big[ (V_{ij} - \overline{V})[ \Theta(\overline{V} - V_{ij}) + \gamma  \Theta(V_{ij} - \overline{V}) ] \Big].
\label{APMgeneral}
\end{eqnarray}
Here, $\Theta(z)$ is the Heavyside function ($\Theta(x>0) =1$ and $\Theta(x<0)=0$) and $\overline{V}$ is an adjustable background value. As the astute reader undoubtedly noticed, the locality constraint imposed by the Kronecker delta in Eq. (\ref{APM'}) has been made explicit in Eq. (\ref{weigh}) by having only intra-community sums for each of the $q$ communities $\{C_{s} \}$. Details of the construction of the weights $V_{ij}$ are provided in \cite{grant6}. Following our more colloquial description here, there are four or five adjustable parameters in Eq. (\ref{weigh}): the resolution parameter $\gamma$, the background value $\overline{V}$, the block size $L$ centered about each pixel (or more general rectangular blocks
of size $L_{x} \times L_{y}$), and the pixel distance $\ell$ over which the pixel interconnection function $V_{ij}$ decays. Once these are set, the earlier community detection algorithm of Section \ref{CD-section} may be applied. The determination of the optimal value(s) of these parameters may be performed using the same procedure outlined in Section \ref{multi}. 

While systems such as the synthetic hierarchical network of Figures \ref{fig:examplehierarchydata} exhibit well defined plateaux in the information theory and other measures, we found more generally that the optimal values of parameters $z$ correspond to local extrema whereby variations in the parameters do not alter the outcome. That is, if $Q$ is a measured quantity of interest (e.g., information theory correlations, Shannon entropy, the energy associated with the given Hamiltonian) then optimal parameters $z$ are found by the requirement that $\nabla_{z} Q  =0$.  These may lead to multiple viable solutions corresponding to very different meaningful partitions. 

In practice, we found that in all but the hardest cases, meaningful solutions are found when arbitrarily setting all parameters to a fixed value and that, similar to Section \ref{multi}, the multi-scale solutions may be found by only varying the resolution parameter $\gamma$. The results of our method are given in Figure (\ref{fig:hardzebradogresults}); these correspond to typical partitions found with the optimal parameter set. The above image analysis ideas may be applied for the detection of the primitive cells in simple Bravais lattices, the inference of domain walls in spin systems, and hierarchical structures in quasi-crystals \cite{grant6}. For a complete classification of contending partitions and, most notably, a deeper understanding of whether the found solutions are meaningful or not, it is useful to survey the canonical finite temperature phase diagram associated with Eq. (\ref{weigh}) when all of the above parameters, including temperature, are varied. In the current context, by ``temperature'', we allude to the finite temperature study of the Hamiltonian of Eq. (\ref{APM'}) either analytically or via a thermal bath associated with, e.g., the acceptance of the moves in the algorithm outlined at the end of Section \ref{CD-section} \cite{grant3,grant6,phases,phases1,phases2}.


\myfig{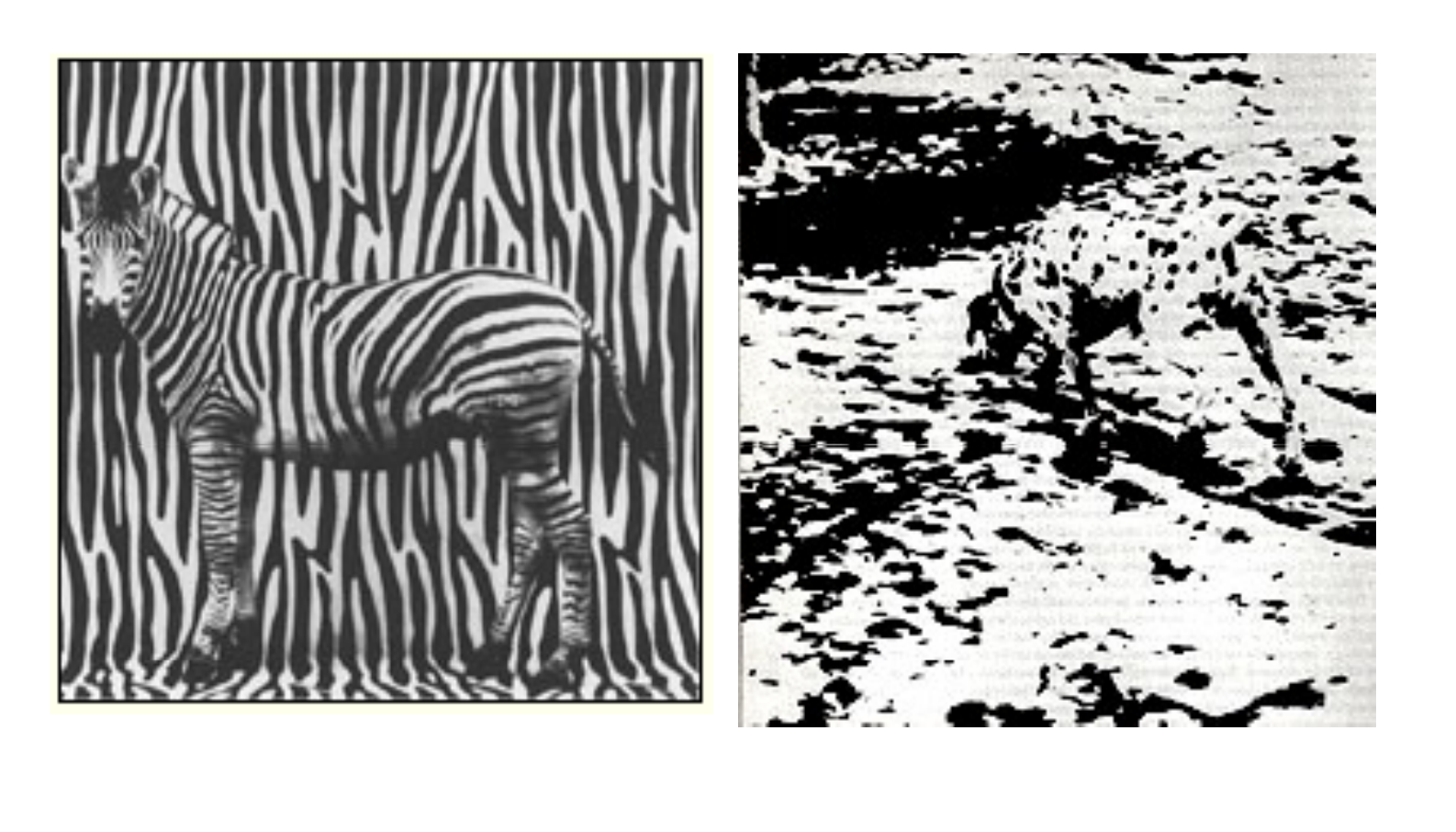}{Examples of the image segmentation
challenges \cite{chal1,chal2}. Left:  The left image is that of  zebra with the a similar``stripe'' background.
Right: The image on the right is that of a dalmatian dog. Most people do not initially recognize the dog
before given clues as to its presence. Once the dog is seen it is
nearly impossible to perceive the image in a meaningless way. }{fig:hardzebradog}{1\linewidth}{}

\myfig{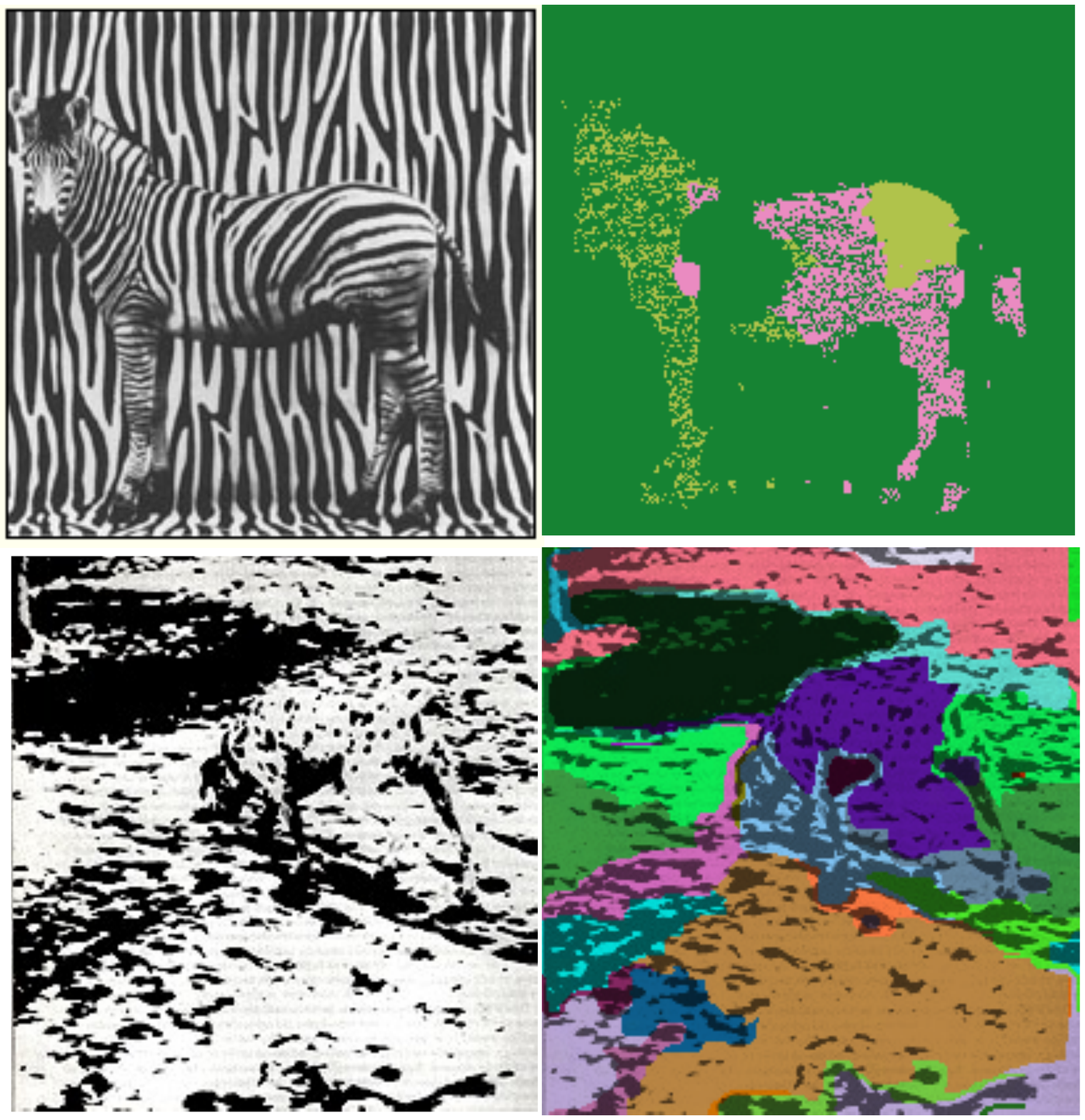}{The application of multiresolution algorithm for the segmentation of the zebra and dalmatian dog images of Figure (\ref{fig:hardzebradog}). The results correspond to typical partitions found with the optimal parameter set. The first and the second rows contain ``camouflages'' of a similar stype. We are able to detect the boundary of the zebra and discern the body and hind legs of the dog albeit with some ``bleeding'' \cite{grant6}.}{fig:hardzebradogresults}{1\linewidth}{}


\section{Community detection phase diagram}
\label{CDPD}
As the bare edge weights and additional parameters setting the values of $V_{ij}$ in the Hamiltonian of Eq. (\ref{weigh}) and temperature are modified, quantities such as the system energy, Shannon entropy, the number of communities, and information theory correlations amongst the found ground states generally attest to the presence of multiple phases. Additional metrics including the ``computational susceptibility'' (the change in the average inter-replica NMI as the number of trials, see Section \ref{CD-section}, is increased \cite{ref:rzmultires,phases,phases1}), the time required for convergence (when attainable), and the ergodic/non-ergodic character (``chaotic'' type feature) of the dynamics all delineate the very same phase diagram boundaries inferred from each of the examined quantities. Information theory measures have been used to
study other specific interesting systems, e.g., \cite{information}. The observed phases in the community detection problem naturally extend to finite temperatures ($T$) when the analysis of the system defined by the Hamiltonian of Eq. (\ref{weigh}) is broadened to include positive temperatures. Finite size systems such as the real networks and images that we discuss cannot exhibit thermodynamic phase transitions and all finite temperature functions are analytic. Nevertheless, practically, sharp changes appear as temperature and other parameters are varied.

Similar to other NP hard \cite{cook} combinatorial optimization problems \cite{hard,mona,gangof3}, three prototypical phases were established in general community detection problems with a distribution of varying community sizes \cite{phases}. Subsequently, these have been beautifully explored in depth in several specific graph types---most notably the so-called ``stochastic block models'', in which a graph has equal size communities e.g., \cite{decelle,sly,NN,darst"} and in other penetrating works, e.g., \cite{steeg,monta1,zhang"}. Earlier signatures of a bona fide transition in stochastic block and power law distributed models \cite{ref:rzlocal,ref:rzmultires} and limits on detectability in the stochastic block model
via the cavity approximation were suggested \cite{ref:ML}. To intuitively highlight the essential character of the prototypical phases with a minimum of jargon, we will colloquially term these the ``easily solvable'', the ``solvable hard'', and the ``unsolvable'' phases. 

In realistic finite yet very large scale systems \cite{phases1,phases2} various results can be established and these may be further examined in various limits. Of course, bona fide transitions formally occur only in the thermodynamic limit. A trivial behavior results in infinite size graphs when the average number of nodes per community is of finite size \cite{phases1,phases2}. As one would expect, typically all community detection problems are either solvable or unsolvable. In NP hard problems, the solvable phase splinters into an ``easy'' and a ``hard phase''. When the edge weights set by $V_{ij}$ are associated with sharp community detection partitions, then finding a natural solution is rather trivial (and nearly all algorithms, not only the Potts model described here, will readily unearth such an answer). On the other hand, if the couplings $V_{ij}$ are sufficiently ``noisy'' so as to be of, effectively, equally the same strength for edges between nodes in the same putative community as for edges linking nodes belonging to different supposed communities, then no well defined community detection solutions exist. Similarly, at sufficiently high temperatures, in most cases, all traces of structures found in the ground state(s) are lost. The most common variant of the community detection problem has been proven to be NP complete \cite{book_comm}.

As in disparate NP problems \cite{gangof3}, it was found that in broad classes of the community deception problem (and in its image segmentation variant) \cite{grant6,phases,phases1,phases2,decelle,NN,steeg,zhang"}, lying between the extremities of the ``easy'' and ``unsolvable'' phases there often exists a ``hard phase'';  in this phase, solutions exist, but due to the plethora of competing states, they may be extremely hard to find. Information theory measures may be used to delineate phase boundaries \cite{grant6,phases,phases1,phases2}. Using information theory correlations and the global Shannon entropy, we show, in Figure (\ref{fig:phasediagramNMI}) and Figure (\ref{fig:phasediagramH}) respectively, the phase diagram associated with the image shown in the upper lefthand side of Figure (\ref{fig:phaseimagebird}). In the solvable phase(s), typically, all partitions produced by parameters that lie in the same basin, lead to qualitatively similar results. Moderate temperature and/or disorder can lead to order by disorder or annealing effects (similar to those found in other systems, e.g., \cite{kirk,villain1980,Henley89,Nussinov04,funnel}). However, at sufficiently high temperatures and/or the introduction of noise about the initial $V_{ij}$ values, the system will be in the unsolvable phase. By carefully studying the system phase diagram and the character and magnitude of the information theory overlaps or thermodynamic functions such as the internal energy and entropy as well as the dynamics, one may assess whether the perceived community detection solutions may be meaningful. When applied to image segmentation, the consistency of this procedure may be inspected visually and intuitively judged sans complicated analysis.


\myfig{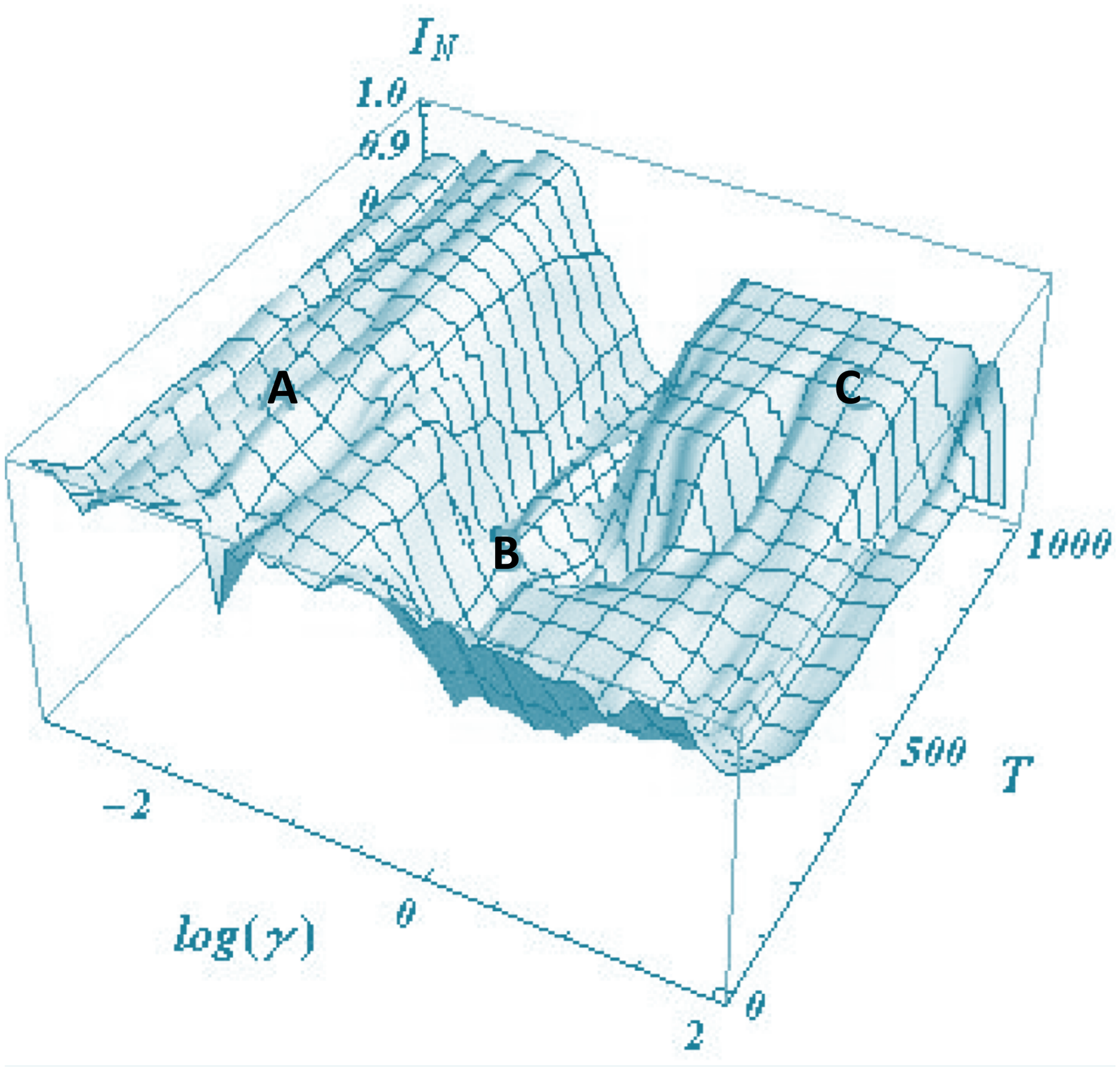}{The normalized mutual information $I_N$ as the function of the resolution $\log(\gamma)$ and temperature $T$ for the ``bird'' image in the upper lefthand panel of Figure (\ref{fig:phaseimagebird}). We mark the ``easy'' phase (where $I_N$ is almost 1 as ``A'', the ``hard'' phase where $I_N$ decreases as ``B'', the ``unsolvable'' phase where $I_N$ forms a plateau whose value is less than 1 as ``C''. The ``easy-hard-unsolvable'' phases will be further confirmed by the corresponding image segmentation results in 
Figure (\ref{fig:phaseimagebird}), as these appear, respectively, in panels A, B, and C therein.)}{fig:phasediagramNMI}{1\linewidth}{}

\myfig{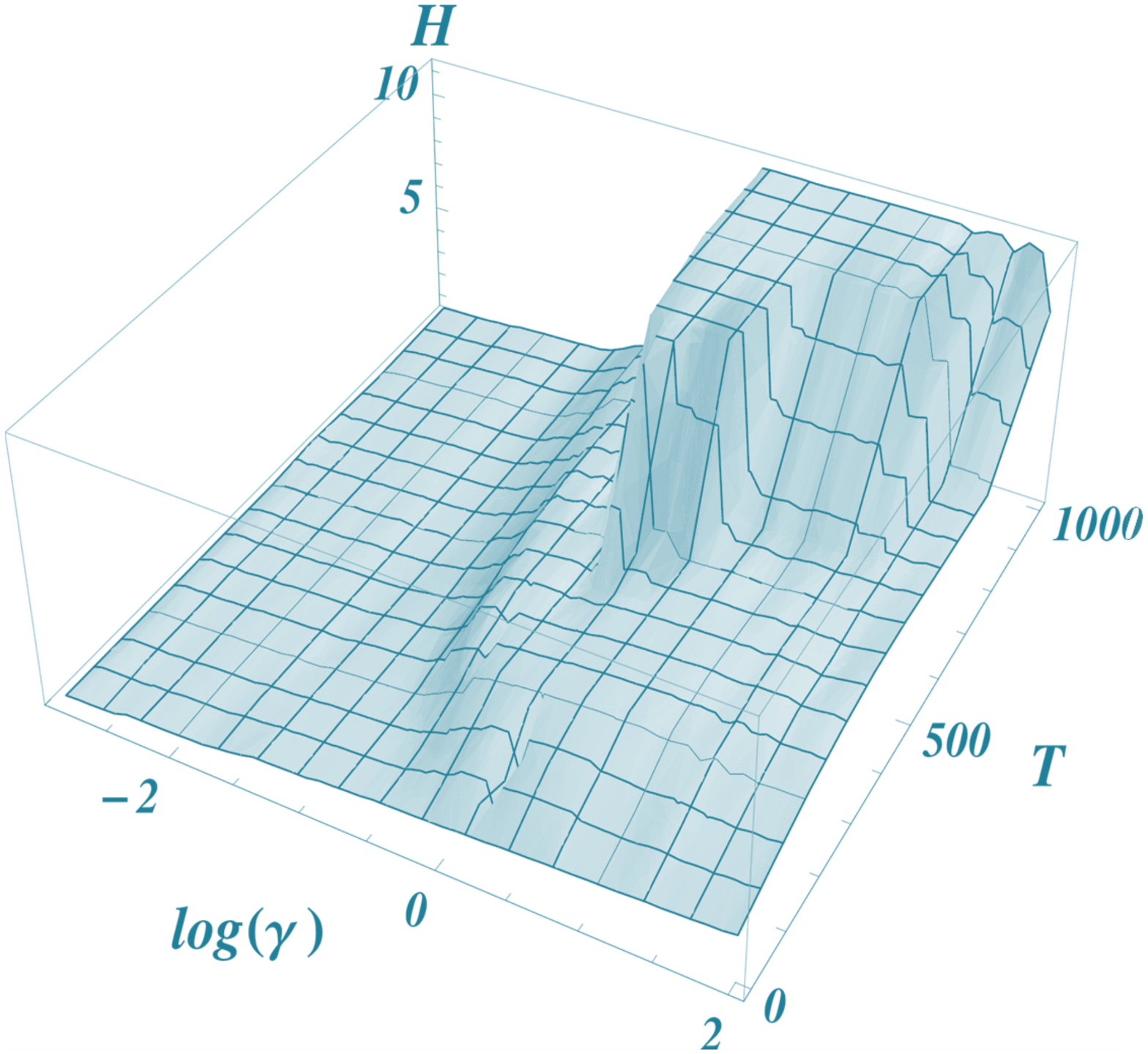}{The Shannon entropy $H$ as the function of the resolution $\log(\gamma)$ and the temperature $T$ for the ``bird'' image in the upper lefthand panel of
 Figure (\ref{fig:phaseimagebird}). The signatures of the three phases ``easy'', ``hard'' and ``unsolvable'' are easily detected in this phase diagram and agree with those ascertained via the normalized mutual information of Figure (\ref{fig:phasediagramNMI}).}{fig:phasediagramH}{1\linewidth}{}



\myfig{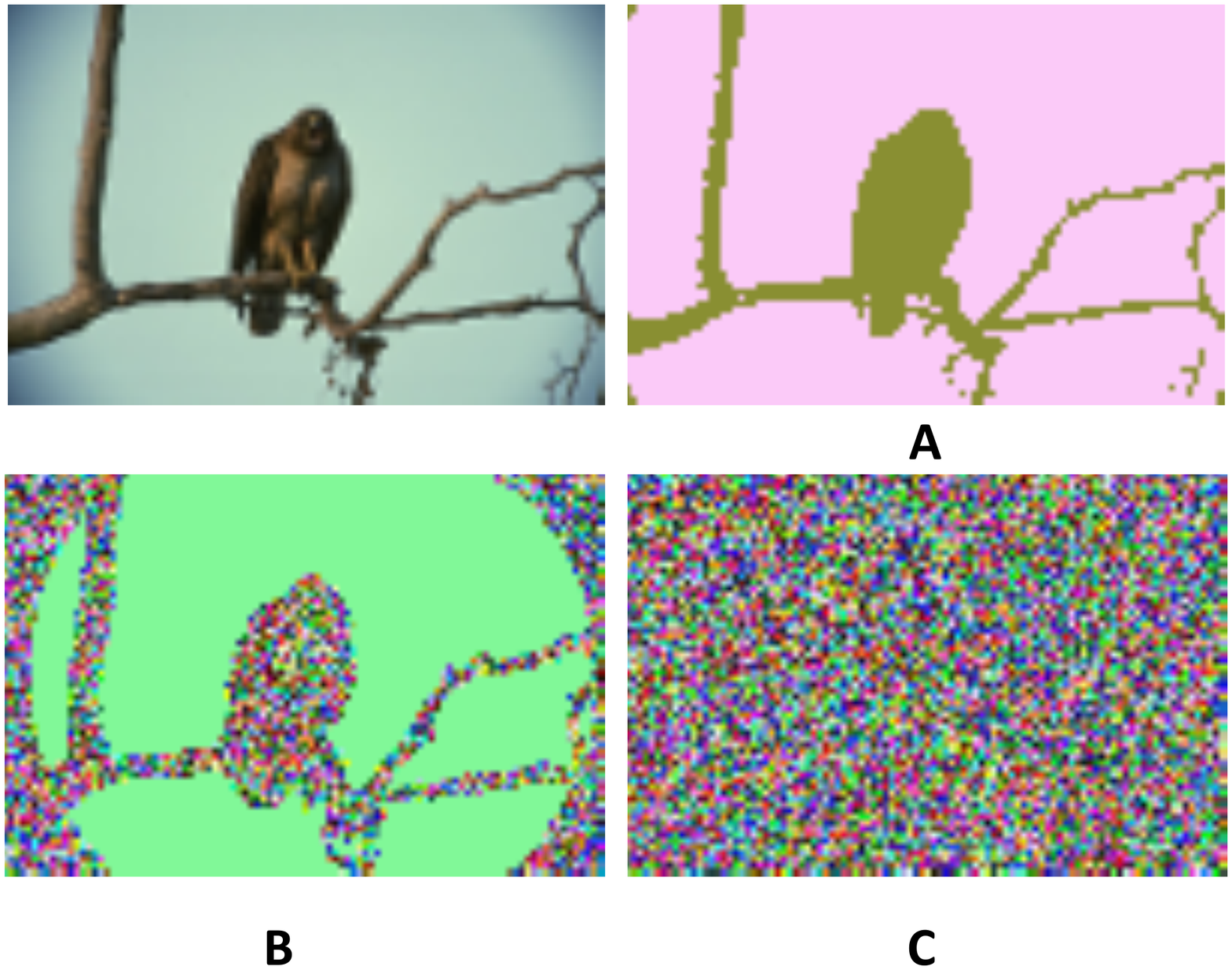}{The image segmentation results of the ``bird'' image. The original image is on the upper left. The other images denoted as ``A'', ``B'', and ``C'' correspond to the image segmentation results with different parameter pairs $(\log (\gamma), T)$ marked in *. Both result A and B are able to distinguish the bird from the ``background''. However in panel B, the bird is composed of lots of small clusters. Result C is unable to detect the bird. Thus, the results shown here demonstrate the corresponding ``easy-hard-unsolvable'' phases in the phase diagram in Figures (\ref{fig:phasediagramNMI},\ref{fig:phasediagramH}). From \cite{grant6}.}{fig:phaseimagebird}{1\linewidth}{}


\begin{figure}
\sidecaption
\includegraphics[width=0.6\columnwidth]{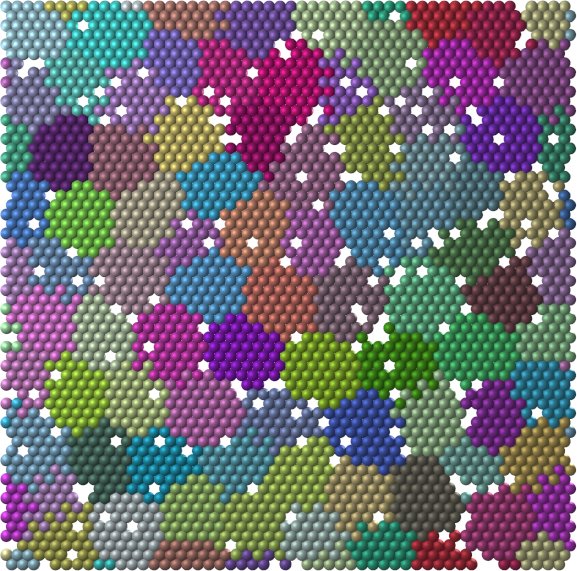}
\caption{A diluted two-dimensional Lennard Jones system with edge weight set equal to the pair interaction energies. 
The ground state of a two dimensional Lennard Jones model is that of a triangular lattice in which the lattice spacing is equal
to the distance at which the Lennard Jones potential attains its minimum. In this figure, the triangular lattice is diluted
by introducing defects in the form static vacancies (denoted by white holes). The found community boundaries are 
intuitively relegated defects lying on the periphery of these domains \cite{grant3}.}
\label{fig:LJnetwork}
\end{figure}

\begin{figure}
\sidecaption
\subfigure[\ Time independent replicas]
  {\includegraphics[width=\widesubfigsize]{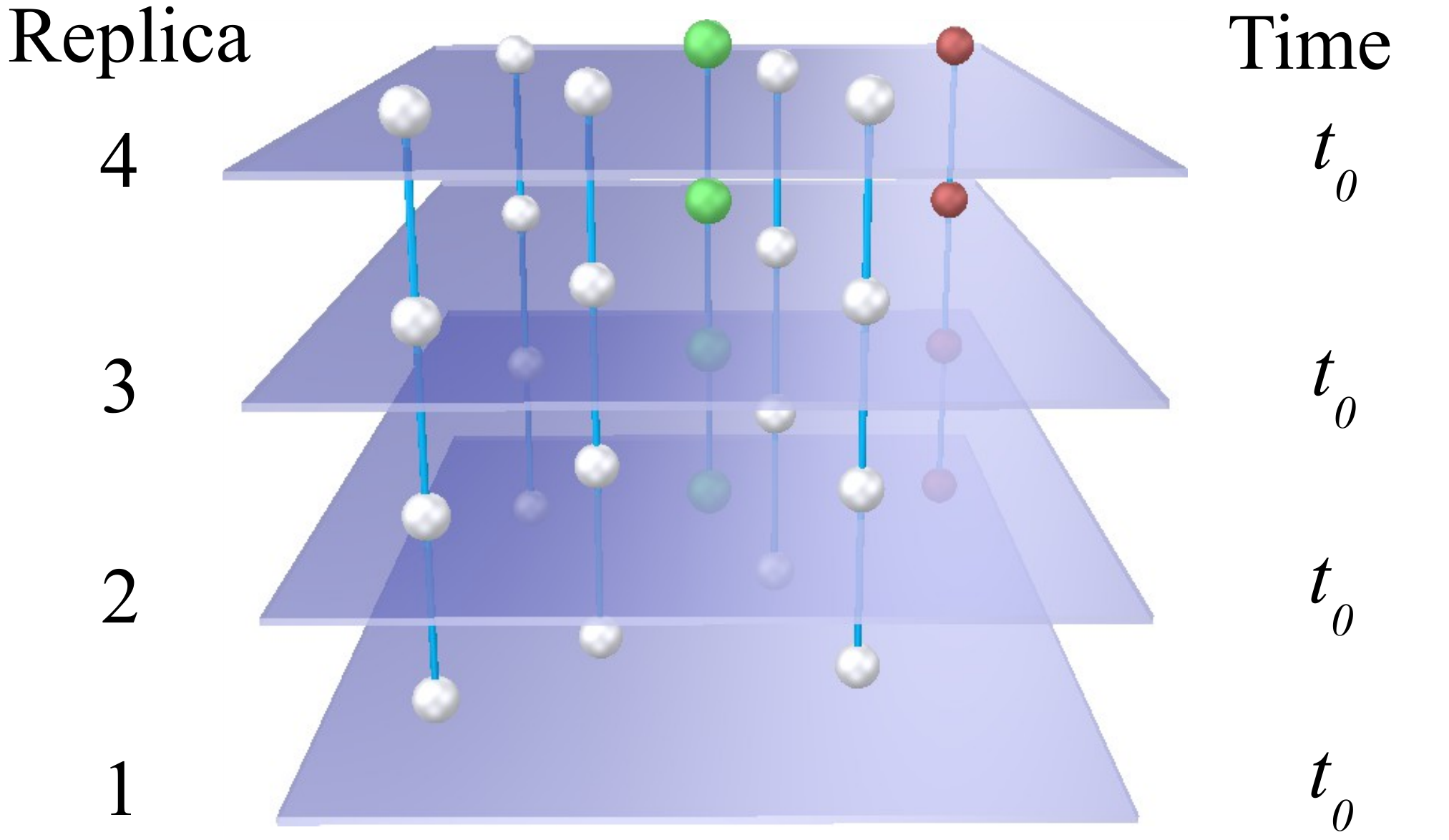}}
\subfigure[\ Time dependent replicas]
  {\includegraphics[width=\widesubfigsize]{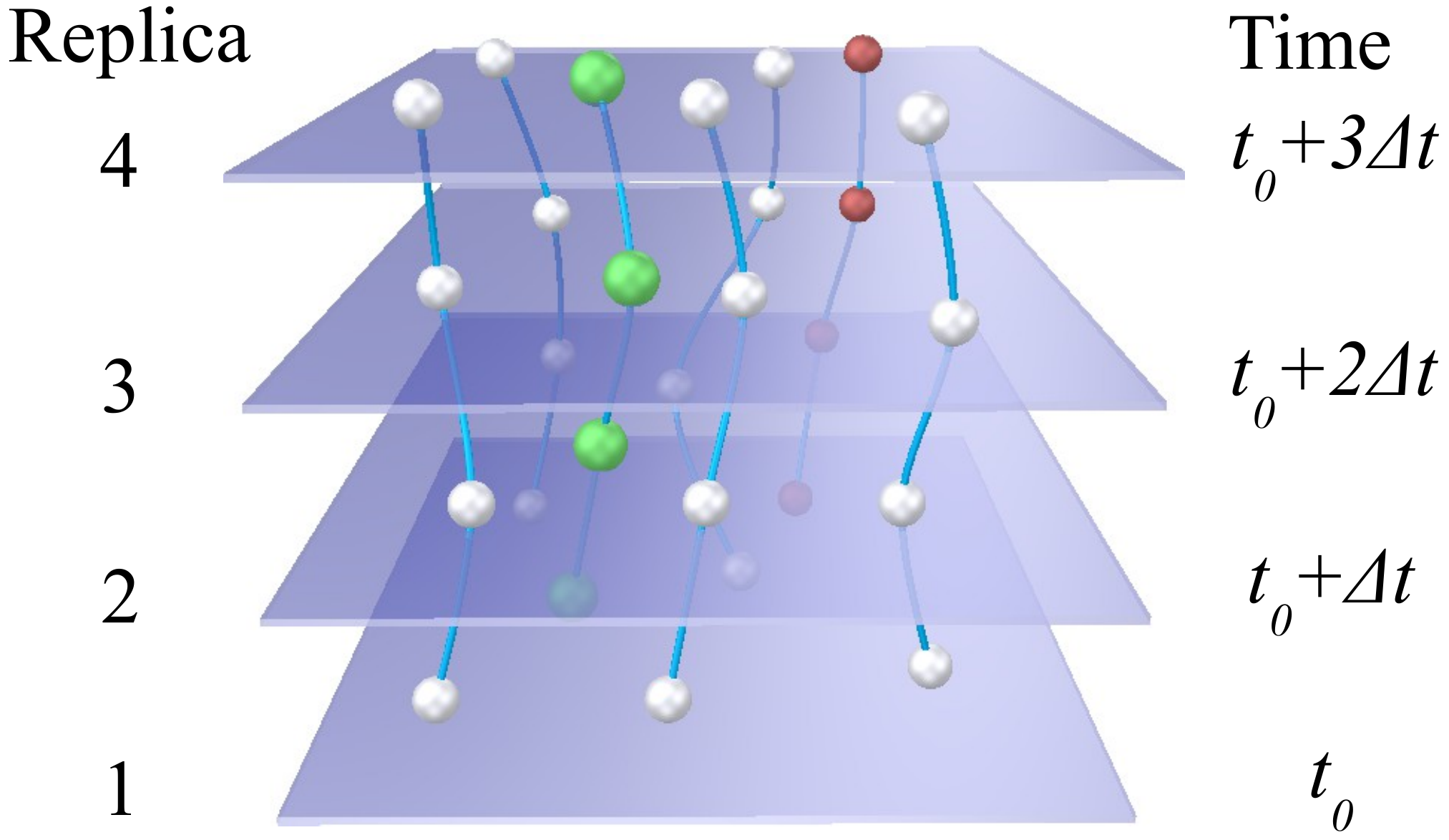}}
\caption{
From \cite{grant3,grant7}. In order to apply the algorithm in \Secref{multi} to complex physical systems, we may generally define two types of replica sets.
Panel (a) depicts a few nodes as they appear for a static system- i.e., one with no time separation between simulation replicas.
Panel (b) depicts a similar set of replicas with each separated by a successive amount of simulation time $\Delta t$. 
In either case, we then generate the replica networks using the potential energy between the atoms as the respective edge 
weights in the network. Consequently, we minimize Eq. (\ref{APM}) using a range of $\gamma$ values in the algorithm described in Section \ref{multi}.}
\label{fig:replicalevels}
\end{figure}

\begin{figure}
\sidecaption
\includegraphics[width=0.6\columnwidth]{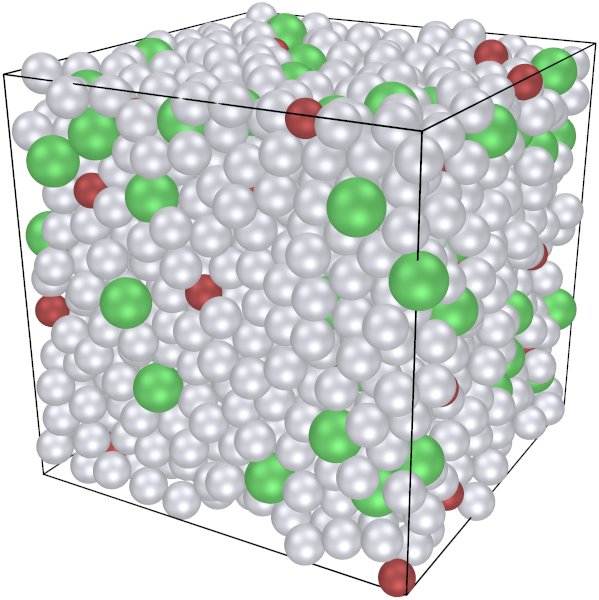}
\caption{From \cite{grant3}. A static snapshot from a molecular dynamics simulation of Al$_{88}$Y$_{7}$Fe$_{5}$ system of 1600 atoms that has been quenched from an initial temperature of 1500 K to 300 K and then allowed to partially equilibrate.
The atoms are Y, Al, and Fe, respectively, in order of increasing diameters. In this figure, the atoms are color coded-- Fe atoms are red and Y atoms are marked green.}
\label{fig:AlYFesystem}
\end{figure}

\begin{figure}
\sidecaption
\includegraphics[width=0.6\columnwidth]{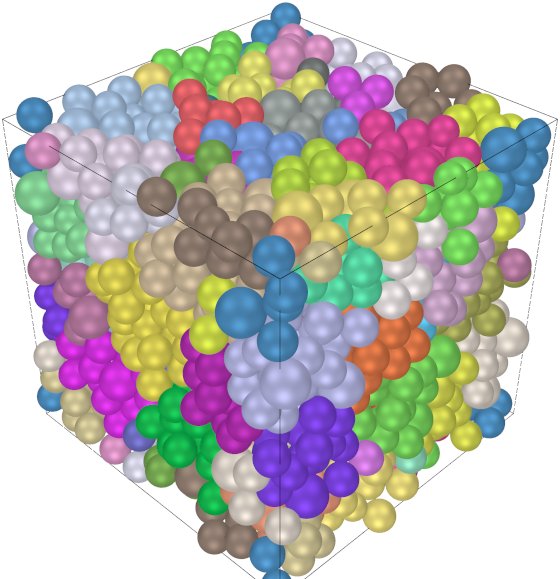}
\caption{The figure shows a static partition of Figure (\ref{fig:AlYFesystem}).
Here, different clusters are identified by individual colors. It is also possible to incorporate overlapping nodes in neighboring clusters to account for the possibility of multiple cluster 
memberships per node, yielding an interlocking system of clusters \cite{grant3}.}
\label{fig:AlYFepartition}
\end{figure}

\section{Casting complex materials and physical systems as networks}

With all of the above preliminaries, we now finally turn to the ultimate data mining objective of this work: that of the important detection of spatial and temporal structure in complex materials and other systems \cite{grant3,grant7,BODP,spatial,holme,emergence,temporal,jack}. This problem shares a common conceptual goal with image segmentation yet is, generally, far more daunting for human examination. Similar to the analysis presented thus far, the approach that we wish to discuss casts physical systems as graphs in space or space-time and then employs the above discussed multi-scale community detection to determine meaningful partitions.

In this case, nodes in the graph code basic physical units of interest (e.g., atoms, electrons, etc.). Multi-particle interactions or experimentally measured correlations in the physical system are then ascribed to edge weights $V_{ij}$ between the nodes (for two-particle interactions or experimentally measured pair correlations \cite{grant3,grant7}), or to three-node triangular weights (for three-particle interactions or correlations) $V_{ijk}$, and so on. Given these static or time-dependent weights, the graph is then (similar to the discussion in earlier sections) partitioned into ``communities'' of nodes (e.g., clusters of atoms) that are more tightly linked to or correlated with each other than with nodes in other clusters \cite{ref:rzlocal}. As in the earlier examples explored in this work, information theory based multi-scale community detection provides both local structural scales (e.g., 
primitive lattice cell, nearest neighbor distance, etc.) as well as global scales (such as correlation lengths) and any other additional intermediate scales if and when these are present. 

The results of this approach for a two-dimensional Lennard-Jones system with vacancies are shown in Figure \ref{fig:LJnetwork}. When the edge weights between nodes are set equal to the Lennard-Jones strength associated with the distance between them, the multi-scale community detection algorithm recognizes both the typical triangular unit cells as well as larger scale domains (communities) in which the vacancy defects tend, on average, to lie on their boundaries. Partitions in which defects tend to aggregate at the domain boundaries is consistent with general expectations for stable domains and is intuitively appealing.


\begin{figure}
\sidecaption
\includegraphics[width=0.79\columnwidth]{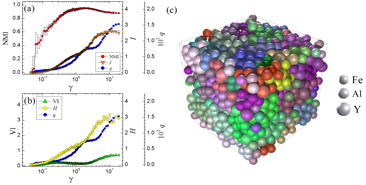}
\caption{The result of the multiscale community detection applied to a ternary glass former at a simulation temperature of $T = 300K$ \cite{grant3,grant7}. Both panels (a) and (b) on the left depict the information theory correlations between the replicas (as described in Section \ref{multi}). In panel (c), each of the communities found is assigned a different color. These structures correspond to the Normalized Mutual Information (NMI) or Variation of Information (VI)  extrema. These well-defined structures contrast sharply with the lack of cohesive features in Figure \ref{fig:ternarydynamicTHigh}.}
\label{fig:ternarydynamicTLow}
\end{figure}
\begin{figure}
\sidecaption
\includegraphics[width=0.79\columnwidth]{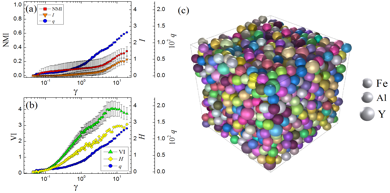}
\caption{The structure of same ternary glass former of Figure \ref{fig:ternarydynamicTHigh} 
at a simulation temperature of $T = 1500K$. Inter-replica information theory correlations are provided in panels (a) and (b). As is evident in panel (c), depict the corresponding lack of structure by significantly higher VI or lower NMI as compared to those in Figure \ref{fig:ternarydynamicTLow}.}
\label{fig:ternarydynamicTHigh}
\end{figure}

\begin{figure}
\sidecaption
\includegraphics[width=0.7\columnwidth]{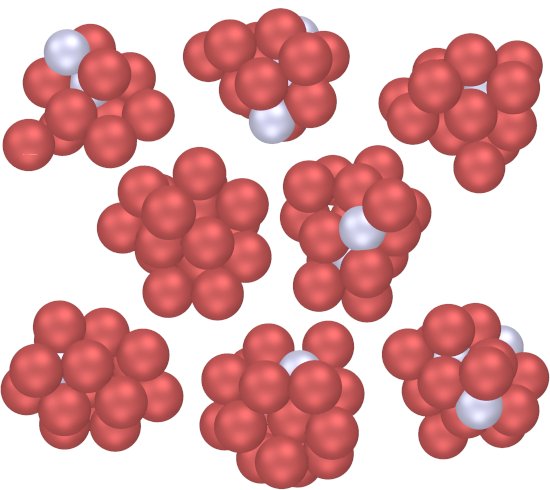}
\caption{From \cite{grant3}. A set optimal clusters found in a low temperature Kob-Andersen system \cite{KA} in which two types of atoms  (color coded 
red and silver) appear.}
\label{fig:LJC}
\end{figure}

As the reader may envisage, the community detection method may be extended to general many-body systems with different types of species (e.g., disparate ion types in metallic glass formers \cite{grant3,grant7}). 
One example is depicted in Figures \ref{fig:replicalevels}, \ref{fig:AlYFesystem}, and \ref{fig:AlYFepartition} corresponding to a ternary system of Al$_{88}$Y$_{7}$Fe$_{5}$based on a molecular dynamics simulation of $1600$ atoms in which edge weights were set by pair potentials is provided in Figs. (\ref{fig:AlYFesystem}, \ref{fig:AlYFepartition}).
As seen in the partition of Figure \ref{fig:AlYFepartition}, for which the inter-replica information theory were extremal and which lies in the solvable phase, below the liquidus temperature (the temperature at which the system is an equilibrium liquid), large clusters were detected. Along similar lines, clusters may be identified across many problems. In Figure (\ref{fig:LJC})
we show typical clusters found in a Kob-Andersen binary system. While for human analysis the complexity of potentially identifying pertinent clusters
may grow dramatically with the number of atom types, for the mutli-resolution analysis there is no such increase. 

In a similar manner, the edge weights can be set by experimentally measured pair correlations. In \cite{grant3}, atomic configurations consistent with the experimentally determined scattering data for quenched Zr$_{80}$Pt$_{20}$ \cite{nakamura,saida,sordelet,wang} were generated \cite{grant3,grant7} using Reverse Monte Carlo methods \cite{mcgreevy,keen}. 

At low temperatures, typically the found structures in all of these cases are far larger than local patterns probed for and detected by current methods \cite{ps,ref:mosayebiCLSGT,ref:berthierCL,ref:karmakarsastry,kl}. Four-point correlations have long been employed to ascertain spatio-temporal scales and the quantify ``dynamical heterogeneities'', e.g.,  \cite{ref:karmakarsastry,4p}.  A long-standing challenge is the identification of structures {\em of general character and scale} in amorphous systems. There is, in fact, a proof that as supercooled liquid falls out of equilibrium to become an amorphous, there must be an accompanying divergent length scale \cite{ref:montanariCL}. 
Methods of characterizing local structures \cite{Sheng,ref:finney,HA,BO} center on a given atom or link; as such, they are restricted from detecting general structures.  Because of the lack of a simple crystalline reference, the structure 
of glasses is notoriously difficult to quantify beyond the very local  scales. In Refs. \cite{grant3,grant7,grant6}, graph weights were determined empirically (potentials in a model system, experimentally measured partial pair density correlations in supercooled fluids, or pixels in a given image)---no theoretical input was invoked as to what the important scales should be or if an exotic order parameter may be concocted. Similarly, in a time dependent analysis for dynamically evolving systems, by employing replicas at different time slices as well as regarding the system as a higher dimensional ``image'' in space-time, using the inter-replica information theory correlations, spatio-temporal patterns were found and time dependent structures were quantified.  In this approach, the data speak for themselves. We remark that notwithstanding the aforementioned difficulties, recently extremely large growth of static structure was observed by far simpler network analysis in certain binary metallic glasses that exhibit crisp icosahedral motifs \cite{grant22}. Similar to the description above, one may likely find other motifs in other systems. The problem is that guessing and hopefully finding pertinent patterns can be extremely challenging to do by conventional analysis.   

\section{Summary}
In this work, we reviewed key features of a statistical-mechanics-based ``community detection'' approach to find pertinent features and structures (both spatial and temporal) in complex systems. In particular, we illustrated how this method may be applied to image segmentation and the analysis of amorphous materials. The demand for automated data mining approaches may become more acute with ever increasingly available data on numerous complex systems. The study of complex materials may be extremely challenging to carry out by current conventional means that rely on guessed patterns, simplified models, or brute force human examination. 

\section{Acknowledgments}
We have benefited from interactions with numerous colleagues. In particular, we would like to thank S. Achilefu, S. Bloch, R. Darst,
S. Fortunato, V. Gudkov, K. F. Kelton, T. Lookman, M. E. J. Newman, S. Nussinov, D. R. Reichman, and P. Sarder 
for numerous discussions and collaboration on some of the problems reviewed in this work and their outgrowths. We are further grateful to support by the NSF under Grants No. DMR-1106293 and DMR-1411229. ZN is indebted to the hospitality and support of the Feinberg foundation for visiting faculty program at the Weizmann Institute.

%
%
%

\end{document}